\documentclass[journal,twocolumn]{IEEEtran}

\usepackage{cite}
\usepackage{amsmath,amssymb,amsfonts}
\usepackage{algorithm,algpseudocode} % 非官方
\usepackage{array}
\usepackage[caption=false,font=normalsize,labelfont=sf,textfont=sf]{subfig}
\usepackage{graphicx}
\usepackage{textcomp}
\usepackage{stfloats}
\usepackage{verbatim}
\usepackage{xcolor, xspace}
\usepackage{booktabs, multirow}
\usepackage{cuted}%%stripsep-3pt
\usepackage{enumitem}
\usepackage{url}
\def\BibTeX{{\rm B\kern-.05em{\sc i\kern-.025em b}\kern-.08em
    T\kern-.1667em\lower.7ex\hbox{E}\kern-.125emX}}
\usepackage{balance}

\newcommand{\Toolname}{\textsc{HF-DGF}}
\newcommand{\tool}{{\sc \Toolname}\xspace}
\newcommand{\toolBold}{{\sc\bfseries \Toolname}\xspace}

\usepackage{caption, subcaption}
\begin{document}

\title{HF-DGF: Hybrid Feedback Guided Directed Grey-box Fuzzing}

% \author{\IEEEauthorblockN{Anonymous Authors}}
\author{
    Guangfa Lyu\IEEEauthorrefmark{1},
    Zhenzhong Cao\IEEEauthorrefmark{2},
    Xiaofei Ren\IEEEauthorrefmark{1},
    Fengyu Wang\IEEEauthorrefmark{1}\IEEEauthorrefmark{3}\IEEEauthorrefmark{4}\\
    \IEEEauthorblockA{
        \IEEEauthorrefmark{1}\textit{School of Software}, \textit{Shandong University}, Jinan, China\\
        \IEEEauthorrefmark{3}\textit{Quan Cheng Laboratory}, Jinan, China\\
        % \IEEEauthorrefmark{1}Email: lvguangfa@mail.sdu.edu.cn\\
        % \IEEEauthorrefmark{3}Email: wangfengyu@sdu.edu.cn\\
        \IEEEauthorrefmark{2}\textit{School of Cyber Science and Engineering}, \textit{Qufu Normal University}, Qufu, China\\
        % \IEEEauthorrefmark{*}Email: caozhzh@qfnu.edu.cn
    }
    \thanks{This work was supported by the Key Project of Quan Cheng Laboratory (QCLZD202304), and the Project of Provincial Laboratory of Shandong(SYS202201).\\
    \IEEEauthorrefmark{4}Corresponding author: Fengyu Wang, Email: wangfengyu@sdu.edu.cn}
}
\maketitle

% 摘要================================================================================
\begin{abstract}
Directed Grey-box Fuzzing (DGF) has emerged as a widely adopted technique for crash reproduction and patch testing, leveraging its capability to precisely navigate toward target locations and exploit vulnerabilities. However, current DGF tools are constrained by insufficient runtime feedback, limiting their efficiency in reaching targets and exploring state spaces.

This study presents \toolBold, a novel directed grey-box fuzzing framework. Its seed scheduling is guided by a hybrid feedback mechanism integrating control-flow distance, value-flow influence score, and slice coverage. To enable precise control-flow distance feedback, we propose a backward-stepping algorithm to calculate basic block-level seed distances on a virtual inter-procedural control-flow graph (ICFG). For effective state space exploration, we introduce value-flow influence and a corresponding metric, the value-flow influence score. Additionally, to mitigate runtime overhead from hybrid feedback, we adopt a novel selective instrumentation strategy.

Evaluations on 41 real-world vulnerabilities show \toolBold outperforms existing tools: it achieves crash reproduction 5.05\(\times\) faster than AFL, 5.79\(\times\) faster than AFLGo, 73.75\(\times\) faster than WindRanger, 2.56\(\times\) faster than DAFL, and 8.45\(\times\) faster than Beacon on average. Notably, when all fuzzers triggered crashes, \toolBold exhibited the lowest code coverage, demonstrating superior directionality and efficiency. It also surpasses AFLGo, WindRanger, DAFL, and Beacon in static analysis efficiency.
\end{abstract}


% 绪论=======================================================================================
\section{Introduction}
\label{sec:introduction}
Grey-box fuzzing \cite{bohme_boosting_2020, fioraldi_weizz_2020, manes_art_2019} has been widely employed in uncovering vulnerabilities in real-world software, including parsers \cite{mathis_parser-directed_2019, mathis_learning_2020}, web browsers \cite{han_codealchemist_2019, wang_superion_2019}, OS kernels \cite{choi_ntfuzz_2021, jeong_razzer_2019, schumilo_kaflhardware-assisted_2017, tan_syzdirect_2023}, and network protocols \cite{gascon_pulsar_2015, liu_state_2022, li_snpsfuzzer_2022}.

For effective testing of pre-determined target locations in large-scale programs, Directed Grey-box Fuzzing (DGF) has been proposed. In contrast to Coverage-guided Grey-box Fuzzing (CGF), which allocates uniform attention across various code sections to maximize overall code coverage, DGF prioritizes efficient navigation towards specified target locations for in-depth testing. This feature has facilitated its extensive adoption in diverse scenarios: identifying regression bugs by focusing on recently modified code \cite{zhu_regression_2021}, testing vulnerability patches by targeting patched code sites \cite{godefroid_automated_2008, godefroid_dart_2005}, reproducing crashes by concentrating on crashing sites \cite{xuan_crash_2015, soltani_search-based_2018}, verifying static analysis reports by designating suspicious locations as targets, and tracking information flow by specifying source-sink pairs \cite{enck_taintdroid_2014}.

% 总结三个指标影响DGF性能
The performance of a directed grey-box fuzzer hinges on three critical dimensions: 
(1) \textit{target convergence efficiency}---the ability to rapidly reach target locations;
(2) \textit{state space exploration effectiveness}---the capability to thoroughly explore the target state space;
(3) \textit{runtime overhead optimization}---the capacity to minimize unnecessary instrumentation costs.

\subsection{Limitations of Existing Approaches}
\subsubsection{Imprecise Path Navigation}
Most existing DGFs rely on coarse-grained seed distance estimations. For example, AFLGo \cite{bohme_directed_2017} determines basic block distances primarily based on function-level distances from the call graph, overlooking intra-function logical discrepancies. Moreover, indirect calls---prevalent in languages like C/C++---are inadequately modeled in call graphs, leading to significant distance calculation errors. These limitations severely degrade navigation precision.

\subsubsection{Inadequate State Space Exploration}
Current DGFs often prioritize proximity to targets while neglecting state space exploration. However, vulnerabilities like buffer overflows impose strict value constraints, requiring not only target reachability but also extensive state exploration. Existing methods lack mechanisms to balance \textit{proximity-driven navigation} and \textit{state-space exploration}, hindering vulnerability triggering efficiency.

\subsubsection{High Runtime Overhead}
To obtain runtime feedback, conventional DGFs employ extensive instrumentation: full instrumentation of all basic blocks for coverage feedback and distance-based instrumentation of only reachable basic blocks. This heavy instrumentation introduces substantial runtime overhead, necessitating strategies to minimize unnecessary instrumentation while preserving feedback effectiveness.

% 我们的解决方案
% 本文介紹了HF-DGF，一種高級的定向灰盒模糊測試方法，旨在應對現有技術所面临的三個關鍵挑戰：精確的路徑導航、有效的目標狀態空間探索以及最小化的运行时性能開銷。HF-DGF的核心是混合反饋驅動的種子調度機制，創新性地集成了三種不同的反饋機制——控制流距離反饋、值流影響分數反饋和切片覆蓋反饋。

% HF-DGF通過細粒度的控制流距离引导模糊器快速到达目标位置，通過新穎的值流影響分数引导模糊器充分探索目标状态空间以触发漏洞，並通過選擇性插桩策略缓解混合反馈所带来的运行时开销優化性能。这三种反馈在种子调度中的协同所产生的reach-then-explore效应，展示了HF-DGF在結合這些反饋以提升模糊測試效率和效果方面的獨特性。

% HF-DGF通過細粒度的控制流距离解決了「如何到達目標」的問題，通過新穎的值流影響分数解決了「到達後如何觸發漏洞」的問題，並通過選擇性插桩策略優化性能，展示了其在結合這些反饋以提升模糊測試效率和效果方面的獨特性。

\subsection{Our Approach: \tool}
This paper presents \tool, an advanced directed grey-box fuzzing framework addressing the above limitations. \tool innovatively integrates three orthogonal feedback mechanisms---control-flow distance, value-flow influence score, and slice coverage---into a hybrid feedback-driven seed scheduling system, complemented by a selective instrumentation strategy to mitigate overhead.
By leveraging fine-grained control-flow distance for efficient target navigation and combining value-flow influence scores (quantifying dependency strength) with slice coverage (quantifying dependency scope) for systematic state space exploration, \tool enables rapid and comprehensive target traversal---a capability overlooked by traditional single-feedback fuzzers. This synergistic integration establishes a "reach-then-explore" paradigm that significantly enhances both the efficiency of target convergence and the effectiveness of vulnerability discovery.

% 最后一段说如何评估的，并粗略的讲一下实现的提升幅度

\subsection{Experimental Validation}
We evaluated \tool using DAFL's benchmark dataset \cite{kim_dafl_2023}, which contains 41 vulnerabilities from the AFLGo benchmark and 9 real-world programs. Key results:
\begin{itemize}[leftmargin=*]
\item \textbf{Rapid Target Convergence}: \tool cuts the time to reach target locations and trigger crashes across all test cases, achieving an average speedup of 2.56–73.75\(\times\) over state-of-the-art baselines.
\item \textbf{Efficient Vulnerability Triggering}: By systematically exploring target state spaces, \tool triggers more vulnerabilities in 24 hours: 1.16\(\times\) more than AFL, 1.12\(\times\) more than AFLGo, 1.45\(\times\) more than WindRanger, 1.14\(\times\) more than DAFL, and 1.56\(\times\) more than Beacon.
\item \textbf{Overhead Optimization}: Using selective instrumentation, \tool reduces code coverage by 80.9\% vs. traditional full-instrumentation fuzzers (AFL/AFLGo/WindRanger), reaching only 54.8\% of DAFL’s coverage while maintaining high vulnerability discovery rates.
\item \textbf{Component Synergy Validation}: Ablation studies show control-flow distance, value-flow influence scores, and slice coverage drive performance gains. Integrating these techniques reproduces vulnerabilities 6.98\(\times\) faster than AFLGo.
\end{itemize}

\subsection{Contributions}
The main contributions of this paper are summarized as follows:
\begin{itemize}[leftmargin=*]
\item \textbf{Novel Framework Design}: We propose \tool, addressing ambiguous navigation, insufficient exploration, and high overhead via integrating three orthogonal feedback mechanisms (control-flow distance, value-flow influence score, slice coverage) into a hybrid hybrid feedback-driven seed scheduling system, enabling a ``reach-then-explore'' paradigm.
\item \textbf{Precise Distance Calculation}: A backward-stepping approach is devised to compute fine-grained basic-block-level seed distances, resolving navigation ambiguity in traditional fuzzers.
\item \textbf{Enhanced State Exploration}: Value-flow influence score (dependency strength) and slice coverage (dependency scope) collaborate to systematically traverse target state spaces, amplifying exploration of unvisited target state spaces.
\item \textbf{Overhead Optimization}: A selective instrumentation strategy is implemented to mitigate runtime overhead inherent to hybrid feedback. This balances feedback richness and performance, ensuring efficiency without compromising exploration thoroughness.
\item \textbf{Comprehensive Validation}: We develop a \tool prototype and conduct extensive experiments. \tool outperforms state-of-the-art fuzzers in both efficiency and effectiveness.
\end{itemize}

% 研究动机===========================================================================================
\section{Motivation}
\label{sec:motivation}
In this section, we illustrate the motivation behind our research by analyzing the key challenges confronting current directed grey-box fuzzers.

% 距离计算
\subsection{Inaccurate control-flow distance calculation}
Existing DGF tools primarily rely on control-flow distance to guide fuzzing, yet inaccuracies in this metric can induce path selection bias. 

Given a target basic block $T_{b}$ within target function $T_{f}$, AFLGo first calculates the function-level distance $d_{f}(f, T_{f})$ as the shortest distance from function $f$ to $T_{f}$ within the call graph. Then, AFLGo computes the basic block-level distance between basic block  $s$ and target basic block $T_{b}$ using Equation \ref{eq:aflgo1}:
\begin{equation}
    \small
    D(s, T_{b})=
        [ \sum_{m \in G(s)} ( d_{b}(s, m) + c \cdot \min_{f \in F(m)} d_{f}(f, T_{f}) )^{-1} ]^{-1}
    \label{eq:aflgo1}
\end{equation}
where $G(s)$ is the set of basic blocks within the control-flow graph of $s$ that call target-reachable functions. $d_{b}(s, m)$ is the distance from basic block $s$ to $m$ (where $m$ invokes a target-reachable function), $F(m)$ represents the target-reachable functions called by basic block $m$, $d_{f}(f, T_{f})$ denotes the function-level distance, and the constant $c$ is typically set to 10 in practice.

The computation of basic block-level distance relies on function-level distance derived from the call graph, which ignores intra-function logic differences. This oversight compromises the accuracy of basic block-level distance metrics. Moreover, when the target location is deep within program logic, errors accumulate through each traversed function, exacerbating the discrepancy between calculated and actual distances.

Other state-of-the-art approaches inherently share similar limitations. For example, WindRanger \cite{du_windranger_2022} calculates distances between deviation basic blocks under the assumption that all functions are equidistant, overlooking internal structural differences. Meanwhile, although Hawkeye enhances proximity measurement by incorporating the number and distribution of call sites derived from static analysis, it still lacks precision in basic block distance calculation due to its ignorance of call site depths within intermediate functions. This gap underscores the necessity of a novel approach to accurately compute control-flow distances.

% (\S \ref{ssec:controlflow}).

% 值流指导
\subsection{Insufficient exploration of target state space}
% 介绍仅依赖控制流的不足，忽略了数据指导而无法满足特定数据条件
It is widely acknowledged that certain vulnerabilities are only triggered under specific data conditions. For instance, a typical buffer overflow vulnerability is more likely to occur when the input seed approaches the boundary of the vulnerable buffer \cite{haller_dowser_2013}. Existing DGF tools like AFLGo primarily focus on control-flow guidance to aid in reaching target locations, yet they often overlook exploration of the target state space---a subset of the program state space inherently linked to target locations. This oversight impedes effective fuzzing from satisfying the specific data conditions required to trigger vulnerabilities.

% 说明现有工作即使用到了数据流，但是利用方法不高效
Data-flow influence feedback can serve as a guiding mechanism for exploring the target state space. However, the underutilization of data-flow information significantly impedes its effectiveness in guiding fuzzing, leading to suboptimal exploration of the target state space. SelectFuzz \cite{luo_selectfuzz_2023} identifies code sections for selective instrumentation by leveraging data dependencies, but fails to fully capitalize on data-flow information for seed scheduling. WindRanger \cite{du_windranger_2022} utilizes taint analysis to collect data-flow information impacting branch constraints, yet its byte-by-byte mutation and feedback collection introduce substantial runtime overhead. These limitations underscore the necessity of introducing value-flow influence as an effective means to explore the target state spaces.

% (\S \ref{ssec:valueflow}).

% 我们的方法=======================================================================================
\section{Overview of \Toolname}
\label{sec:approach1}
In this section, we present a brief overview of \tool, with particular emphasis on clarifying its core mechanism: path selection.

\subsection{Structure and Workflow}
\label{ssec:overview}
\begin{figure}[ht]
\centering
\includegraphics[width=\columnwidth]{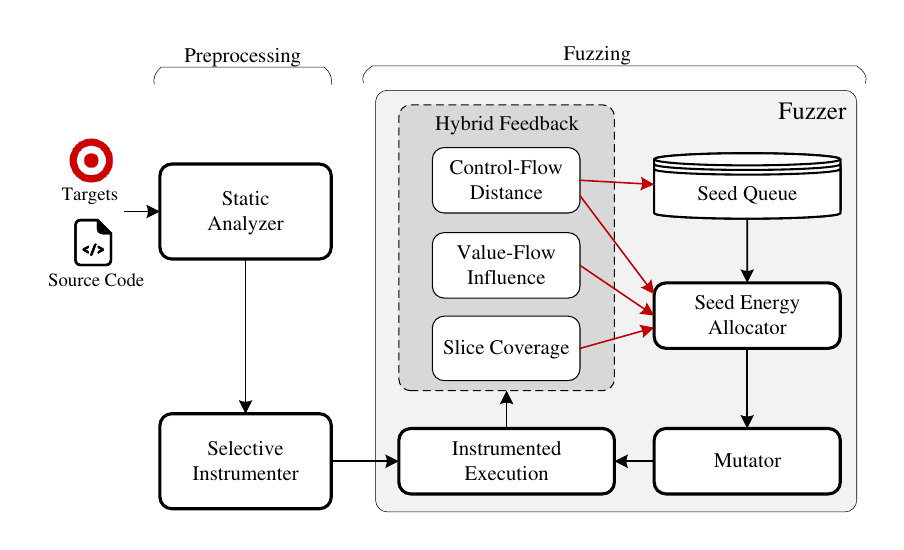}
\caption{Overview of \Toolname.}
\label{fig:arch}
\end{figure}

The overall architecture of \tool is depicted in Fig.~\ref{fig:arch}, which consists of three core components: static analyzer, selective instrumenter, and fuzzer. These components collaborate to enable hybrid feedback-driven fuzzing, and their workflow is structured as follows.

First, the static analyzer performs in-depth analysis on the target program’s source code. It constructs function call graphs, control-flow graphs (CFGs), and value-flow graphs (VFG) to model execution logic and data dependencies. For target-reachable basic blocks, it calculates two key metrics at the basic-block level: control-flow distance (measuring structural proximity to the target in the CFG) and value-flow influence score (quantifying data-flow impact on the target via the VFG). Additionally, it slices these target-reachable basic blocks and identifies boundary basic blocks (those with no sliced successors), which together define the scope for selective instrumentation to ensure only critical regions are instrumented.

Next, the selective instrumenter injects code into the target program based on the static analysis results. Specifically, it instruments sliced basic blocks to obtain slice coverage feedback, boundary basic blocks to capture distance feedback, and blocks with value-flow influence scores to generate value-flow influence feedback. This targeted instrumentation minimizes overhead while ensuring feedback directly reflects the target context.

Finally, the fuzzer executes the instrumented program and leverages runtime feedback for seed scheduling and energy allocation. Seed prioritization is determined by distance to the target (closer seeds get higher priority), and seed energy (number of mutations) is computed by integrating slice coverage, control-flow distance, and value-flow influence. Seeds with higher slice coverage, shorter distances, and stronger value-flow influence receive more mutations, directing fuzzing efforts toward high-potential paths and accelerating vulnerability discovery.

In summary, the workflow of \tool commences with static analysis to map the program structure and delineate the instrumentation scope, proceeds to selective instrumentation for the collection of targeted feedback, and culminates in feedback-driven fuzzing that prioritizes seeds and strategically allocates mutation energy. 

\subsection{Path Selection}
\label{ssec:pathselection}
To elucidate the underlying rationale of the \tool system, we exemplify how seed prioritization and energy allocation influence path selection.

\begin{figure}[ht]
\centering
\includegraphics[width=\columnwidth]{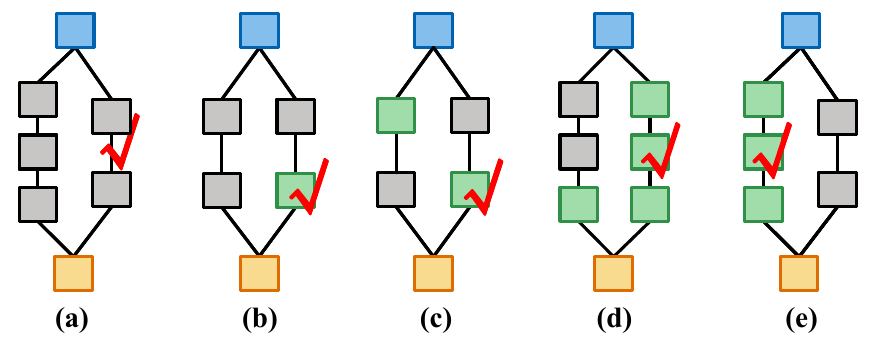}
\caption{Path Selection in \Toolname.}
\label{fig:note}
\end{figure}

As depicted by the CFGs in Fig.~\ref{fig:note}, blue blocks signify entry points, yellow blocks indicate crash-trigger points, and green blocks represent basic blocks that directly or indirectly impact the target value.
In Fig.~\ref{fig:note}a, two paths exist, differing only in distance with no alterations to the target value. Consequently, \tool gives precedence to the shorter path.
In Fig.~\ref{fig:note}b, despite both paths reaching the target at an equal distance, the right-hand path contains a basic block that modifies the target value. Thus, \tool favors the right-hand path.
Fig.~\ref{fig:note}c shows two paths, both modifying the target value. However, the basic block influencing the target value on the right-hand path is nearer, resulting in a more direct influence. Hence, \tool exhibits a preference for the right-hand path.
In Fig.~\ref{fig:note}d, both paths include basic blocks modifying the target value. Yet, the right-hand path has more such blocks, leading to a more extensive value-flow influence and a more thorough exploration of the target state space. Therefore, \tool inclines towards the right-hand path.
For the scenarios in Fig.~\ref{fig:note}e, where paths vary in both distance to the target location and target-value modifications.
% , the preferred path is determined according to the methodology detailed in \S \ref{ssec:scheduling}. 
Initially, \tool explores the shorter right-hand path. But since it doesn't involve value modifications, \tool allocates less energy to it. When it reaches the left-hand path, which has a stronger value-flow influence, \tool assigns more energy to the seed, enabling comprehensive exploration of the target state space.

\section{Design of \Toolname}\label{sec:approach2}

The fuzzing design of \tool hinges on integrating three key feedbacks: control-flow distance, value-flow influence score, and slice coverage. This section first details the calculations for control-flow distance and value-flow influence score. It then presents the selective instrumentation strategy to acquire these three feedbacks. Finally, it explains how the multi-dimensional feedback drives seed prioritization and energy allocation in fuzzing.

% 距离计算
\subsection{Control-flow distance}
\label{ssec:controlflow}
In directed grey-box fuzzing, seed distance feedback is critical for steering the exploration towards target locations. While our approach, similar to AFLGo, computes distances at the basic block level via static analysis and determines seed distances during fuzzing, we innovate by constructing a virtual ICFG and designing a backward-stepping distance calculation algorithm to compute control-flow distances of basic blocks rapidly and precisely. The following subsections detail our methodology in \tool.

\subsubsection*{\textbf{Basic-Block Distance}}
Traditional function-level distance calculation based on call graphs is coarse-grained. By treating function calls as basic block transitions and unwrapping functions to the basic-block level, we derive fine-grained basic-block distances, defined as the shortest path length from a basic block to the target within an ICFG.

\begin{figure}[!t]
\centering
\includegraphics[width=\columnwidth]{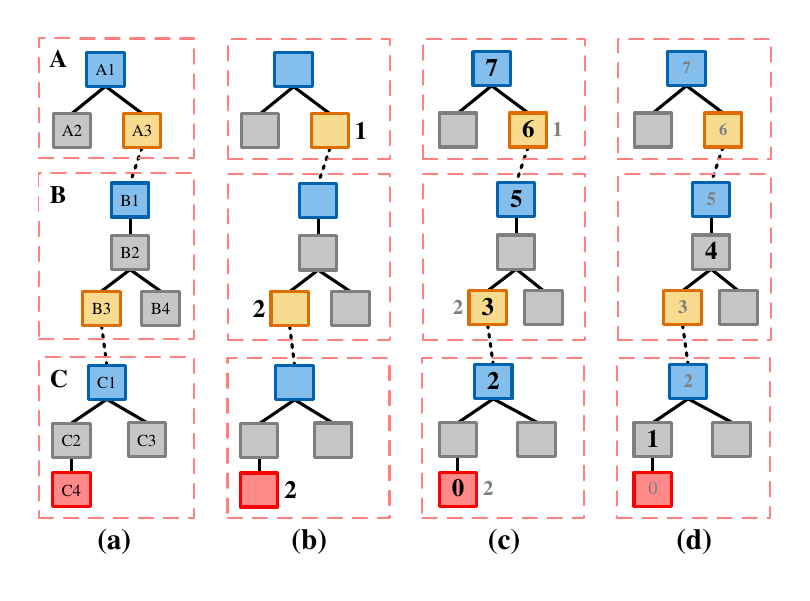}
\caption{Basic-block Distance Calculation in \Toolname.}
\label{fig:distcalc}
\end{figure}

However, constructing a comprehensive path-sensitive ICFG is impractical due to its scale, so we adopt a backward-stepping strategy for basic-block distance calculation in a virtual ICFG. This strategy decomposes the calculation into four manageable steps for accuracy, which we illustrate using Fig.~\ref{fig:distcalc}.

First, we identify functions reachable to the target location via function call sequences (e.g., functions \emph{A}, \emph{B}, \emph{C} marked in Fig.~\ref{fig:distcalc}a) using breadth-first traversal on the function call graph, expanding the scope with Andersen-style pointer analysis \cite{andersen_program_1994} to handle indirect calls. During traversal, we record call-site basic blocks (\textbf{yellow} in Fig.~\ref{fig:distcalc}) invoking these reachable functions.

% First, we identify functions reachable to the target location through the sequence of function calls (e.g., functions \emph{A}, \emph{B}, and \emph{C} marked in Fig.~\ref{fig:distcalc}a). We conduct a breadth-first traversal on the function call graph for this purpose. To account for functions called indirectly through pointers, we use Andersen-style pointer analysis \cite{andersen_program_1994} to expand the scope. During traversal, we record the call-site basic blocks (\textbf{yellow} blocks in Fig.~\ref{fig:distcalc}) that call these reachable functions.

Next, we compute the depth of each call-site basic block as the shortest path length from the entry basic block (\textbf{blue} in Fig.~\ref{fig:distcalc}) within the control-flow graph. For example, in Fig.~\ref{fig:distcalc}b, the depths of \emph{A3}, \emph{B3}, and \emph{C4} are 1, 2, and 2, respectively.

% Next, we calculate the depth of call-site basic blocks. The depth of a basic block is defined as the shortest path length from the entry basic block (\textbf{blue} blocks in Fig.~\ref{fig:distcalc}) within the control-flow graph to that basic block. For example, in Fig.~\ref{fig:distcalc}b, the depths of basic blocks \emph{A3}, \emph{B3}, and \emph{C4} are 1, 2, and 2 respectively.

In the third step, we calculate distances from both call-site and entry basic blocks to the target. For a call-site basic block, its distance is the called function's entry block distance plus 1 (choosing the minimum if multiple functions are invoked). For an entry basic block, its distance is the sum of the depth and distance of the call-site basic block within the same function (taking the minimum sum if multiple call sites exist).
Given the interdependency of these calculations, we follow Algorithm~\ref{algo:dist}, computing distances for blocks \emph{C1}, \emph{B3}, \emph{B1}, \emph{A3}, \emph{A1} sequentially, as shown in Fig.~\ref{fig:distcalc}c.

% In the third step, we calculate the distances from both call-site and entry basic blocks to the target basic block. For a call-site basic block, its distance equals that of the called function's entry basic block plus 1. When multiple functions are invoked at a call-site, we choose the minimum distance. For an entry basic block, its distance is the sum of the depth and distance of the call-site basic block within the same function. When multiple call-site basic blocks exist, we take the minimum sum. Given the interdependent of the two distance calculations, we adhere to the steps presented in Algorithm~\ref{algo:dist}. As illustrated in Fig.~\ref{fig:distcalc}c, we compute the distances of basic blocks \emph{C1}, \emph{B3}, \emph{B1}, \emph{A3}, and \emph{A1} in sequence.

Finally, for the control-flow graph of each function reachable to the target location (i.e., functions capable of reaching the target via call graph traversal), we perform a reverse breadth-first traversal from its call-site basic block, tracking the number of traversed layers. The distance of each traversed block is the call-site block's pre-determined distance plus the layer count. For instance, in Fig.~\ref{fig:distcalc}d, the distances of \emph{C2} and \emph{B2} are 1 and 4, respectively.

More formally, the basic-block distance is defined by Equation \ref{eq:bbdistance}.
\begin{equation}
\small
% \footnotesize
\begin{split}
    & Distance(bb, t) = \\
    &\begin{cases}
        \quad 0                                                           & bb = t \\
        \min\limits_{f \in CF(bb)}  Distance(Entry(f), t) + 1          & bb \in CS(bb) \\
        \min\limits_{cs \in CS(bb)} [ Depth(bb) + Distance(cs, t) ]    & bb \in Entries \\
        \min\limits_{cs \in CS(bb)} [ dbb(bb, cs) + Distance(cs, t) ]  & bb \in Reach(t) \\
    \end{cases}
\end{split}
\label{eq:bbdistance}
\end{equation}
where, $dbb(bb, cs)$ is the shortest length from basic block $bb$ to $cs$ within the control-flow graph;
$call-sites$ represents all call-site basic blocks;
$Entries$ denotes the entry basic blocks of all sliced functions;
$Reach(t)$ contains the basic blocks from which the target basic block $t$ is reachable along the control-flow;
$CF(bb)$ signifies the functions called by the \verb|call| instructions within the basic block $bb$;
and $CS(bb)$ refers to all call-site basic blocks within the function in which $bb$ resides.

\begin{algorithm}[ht]
\caption{Backward-Steping Distance Calculation.}
\begin{algorithmic}[1]
\Require $target\_bb$, $depth$
\Ensure $distance$
\State $target\_function \gets \Call{GetFunction}{target\_bb}$
\State $distance[target\_bb] \gets 0$
\State $distance[target\_function] \gets 0$
\State $function\_queue \gets \{target\_function\}$
\While{$function\_queue$ is not empty}
    \State $callee \gets function\_queue$.dequeue()
    \ForAll{$callsite$ in \Call{GetCallSites}{callee}}
        \State $caller \gets \Call{GetFunction}{callsite}$
        \State $dist \gets distance[callee] + 1$
        \If{$distance[callsite] > dist$}
            \State $distance[callsite] \gets dist$
        \EndIf
        \State $dist \gets distance[callsite] + depth[callsite]$
        \If{$distance[caller] > dist$}
            \State $distance[caller] \gets dist$
            \State $function\_queue$.enqueue($caller$)
        \EndIf
    \EndFor
\EndWhile
\end{algorithmic}
\label{algo:dist}
\end{algorithm}

When multiple target locations exist, we compute the distance from a basic block to each target independently and then use the \textit{harmonic mean} to derive the block's final distance, as defined in Equation~\ref{eq:dist_bb}. This approach ensures the distance metric accounts for proximity to all targets.

\begin{equation}
    \small
    Distance(bb, T) = \left[\sum_{t \in T} Distance(bb, t)^{-1} \right]^{-1}
    \label{eq:dist_bb}
\end{equation}

\subsubsection*{\textbf{Seed Distance}}
Seed distance measures the proximity of a seed's execution path to the target location. Unlike AFLGo \cite{li_directed_2023} and HawkEye \cite{chen_hawkeye_2018}, which compute this metric using all basic blocks, \tool innovatively averages the precomputed control-flow distances of boundary basic blocks. For a target set $T$, the distance of seed $s$ is defined as:
\begin{equation}
    \small
    Distance(s, T) = \frac{ \sum_{bb \in \Phi(s) } Distance(bb, T)}{|\Phi(s)|}
    \label{eq:dist_s}
\end{equation}
where $\textit{Distance}(bb, T)$ represents the precomputed control-flow distance from basic block $bb$ to $T$, and $\Phi(s)$ denotes the set of boundary blocks in $s$'s execution trace. 

This seed distance calculation method offers two key advantages. First, by focusing on boundary basic blocks, it minimizes instrumentation and feedback collection overhead by eliminating redundant monitoring of non-critical code. Second, by using the boundaries of reachable code as the metric, it enables precise seed distance measurement---outperforming traditional all-block averaging methods in both efficiency and accuracy.

% 值流影响反馈
\subsection{Value-flow influence}\label{ssec:valueflow}
Data-sensitive vulnerabilities (e.g., buffer overflows) depend on specific target data states, not just target location reachability. To tackle this, \tool introduces value-flow influence and a novel fitness metric: the value-flow influence score. Unlike DAFL, which uses data-flow analysis to slice basic blocks and compute control-flow distances, \tool employs the value-flow influence score to quantify a seed’s ability to explore target state spaces. The score aggregates influence contributions from all impactful basic blocks along the execution path, inherently prioritizing multi-stage target data manipulation paths. This accumulative scoring strategy enables systematic exploration of complex state spaces, improving the detection of data-sensitive vulnerabilities by guiding fuzzing efforts toward critical data dependencies.

\subsubsection*{\textbf{Value-flow Influence}}
In program analysis, value flow \cite{steffen_value_1990} refers to the transfer of variable values during program execution through assignments and \textit{def-use} chains. Previous studies \cite{cherem_practical_2007, sui_svf_2016} have demonstrated that analyzing program value flow can effectively detect potential vulnerabilities. This underlines the significance of integrating value-flow analysis when modeling data modification and transfer in programs.

Inspired by these findings, we use value flow to guide fuzzing for efficient exploration of the program's target state space. Basic blocks directly or indirectly modifying target data have a value-flow influence. We assign distinct weights, called value-flow influence scores, to these basic blocks. Blocks nearer the target location get higher scores. \tool tracks and accumulates the value-flow influence scores of covered basic blocks for one seed, which becomes the seed's value-flow influence score. A higher score implies the seed is more beneficial for exploring the target state space.

\subsubsection*{\textbf{Value-flow Influence Score of Seed}}
We compute a seed's value-flow influence score in two phases. First, in static analysis, we obtain basic-block-level value-flow influence scores. Then, during fuzzing, we determine the final seed-level score.

The value-flow influence score is derived from the value-flow graph. To identify complex pointer references and structure sub-fields, we enhance the value-flow graph using field-sensitive Anderson-style pointer analysis. In this graph, instructions modifying target data are nodes, and their influence on target data depends on their proximity to it. Thus, we initially calculate the distance from each instruction to the target data using the value-flow graph. For a given target data $t$, we define the value-flow distance of instruction $i$ as:
\begin{equation}
    % \footnotesize
    \small
    \textit{VFD}(i, t) = \begin{cases}
        undefined                                            & i \notin Reach(t) \\
        \qquad 0                                             & i = t \\ 
        \min\limits_{si \in \textit{Succ}(i, t) } \textit{VFD}(si, t) + 1      & i \in Reach(t)
    \end{cases}
\end{equation}

For scenarios with multiple targets, a single instruction can impact multiple data points. To measure an instruction's influence on a multi-target set $T$, we define the instruction-level value-flow influence ($\textit{VFI}(i,T)$) as:
\begin{equation}
    \small
    \textit{VFI}(i, T) = \frac{ \sum_{v \in V(T)}  \left[\max{\textit{VFD}} -  \textit{VFD}(i, v)\right]}{|V(T)|} 
\end{equation}
where $V(T)$ consists of all targets within the multi-target set $T$, and $\textit{VFD}(i, v)$ denotes the value-flow distance from instruction $i$ to a specific target $v$. Essentially, instructions closer to the targets are assigned a higher value-flow influence.

Next, we calculate the basic-block-level value-flow influence as the \textit{minimum} of all instruction-level value-flow influences within the block. For a basic block $bb$ with a set of instructions $ins$, we define its value-flow influence $\textit{VFB}$ as:
\begin{equation}
    \small
    \textit{VFB}(bb, T) = \min_{i \in ins} \textit{VFI}(i, T)
\end{equation}

The seed-level value-flow influence score $\textit{VFS}$ is computed during the fuzzing process. Let $\Phi(s)$ represent the execution trace of seed $s$, encompassing executed basic blocks with value-flow influence. We define the seed's value-flow influence score $\textit{VFS}$ as follows:
\begin{equation}
    \small
    \textit{VFS}(s, T) = \sum_{bb \in \Phi(s) } \textit{VFB}(bb, T)
\end{equation}
The seed-level value-flow influence score is the cumulative value-flow influences of the basic blocks in its execution trace.

During fuzzing, we track the maximum ($\max\textit{VFS}$) and minimum ($\min \textit{VFS}$) value-flow influence scores  of all seeds for final-score normalization. We define the normalized value-flow influence score $\tilde{\textit{VFS}}(s, T)$ of seed $s$ as follows:
\begin{equation}
    \small
    \tilde{\textit{VFS}}(s, T) = \frac{ \textit{VFS}(s, T) - \min \textit{VFS} }{ \max \textit{VFS} - \min \textit{VFS} }
\end{equation}

% 种子调度
\subsection{Seed Scheduling}
\label{ssec:scheduling}
\tool employs multi-dimensional feedback metrics, specifically control-flow distance, value-flow influence, and slice coverage, to direct seed prioritization and energy allocation in the fuzzing process. The following is a detailed explanation of this mechanism.

\subsubsection*{\textbf{Seed Prioritization}}
Drawing on AFL \cite{zalewski_american_2017}, we implement a circular queue as the seed pool. New seeds are prioritized by their control-flow distances upon insertion, enabling \tool to select seeds closer to the target earlier in the fuzzing process. To optimize performance, \tool maintains pointers to every 100th seed in the queue, which streamlines queue maintenance and reduces runtime overhead.

% 能量调度
\subsubsection*{\textbf{Energy Allocation}}
DGF generally determines the mutation count for each seed based on its allocated energy. To generate more mutants from promising seeds, \tool allocates more energy to seeds that are closer to the target, have higher value-flow influence scores, and exhibit greater slice coverage.

To address potential local optima, \tool employs the simulated annealing algorithm independently on both control-flow distance and value-flow influence score. For control-flow distance, \tool adapts AFLGo's exponential cooling schedule $T_{\exp}$. Concurrently, for the value-flow influence score, we introduce a novel annealing-based power schedule:
\begin{equation}
    P_{\Toolname}(s, T) = 2^{10 \cdot \tilde{\textit{VFS}}(s, T) \cdot (1 - T_{\exp}) + 0.5 T_{\exp} - 5}
\end{equation}
where $\tilde{\textit{VFS}}(s, T)$ denotes the normalized value-flow influence score of seed $s$ relative to target $T$, and $T_{\exp}$ is the shared cooling schedule from AFLGo's distance-based annealing. 

The final energy allocated to a certain seed is determined by integrating three dimensions: $P_{AFL}(s, T)$ (the energy assigned by AFL with slice coverage), $P_{AFLGo}(s, T)$ (the energy from AFLGo's control-flow distance metric), and $P_{\tool}(s, T)$ (the energy computed via \tool's value-flow influence score). The integrated energy $P(s, T)$ is calculated as the product of these components:
\begin{equation}
    \small
    P(s, T) = P_{\textit{AFL}}(s, T) \cdot P_{\textit{AFLGo}}(s, T)  \cdot  P_{\tool}(s, T)
\end{equation}
This integration enables a comprehensive prioritization strategy that balances traditional coverage metrics, AFLGo's control-flow distance, and novel value-flow analysis.

% 选择性插桩
\subsection{Selective Instrumentation}
\label{ssec:instrumentation}
DGF enhances fuzzing by instrumenting the target program for runtime feedback, though instrumentation inevitably introduces runtime overhead---thus, balancing detailed feedback with overhead reduction is critical. \tool adopts a non-one-size-fits-all strategy: we design three customized instrumentation approaches for distinct feedback dimensions. For coverage, we instrument only sliced basic blocks; for control-flow distance, we target boundary basic blocks; for value-flow influence, we instrument all blocks with value-flow scores. This strategy embodies a core philosophy: acquiring essential information at the lowest cost from code points with the highest information value, minimizing performance degradation while offering deep insights into program behavior across coverage, control-flow distance, and value-flow influence.

% DGF enhances fuzzing by instrumenting the target program for runtime feedback. However, instrumentation unavoidably raises the target program's runtime overhead. Thus, balancing detailed feedback acquisition with overhead reduction is essential. In \tool, we employ a selective instrumentation strategy. This strategy targets only critical code sections, minimizing performance degradation while offering in-depth insights into program behavior. To gather effective runtime feedback, this strategy encompasses three dimensions: coverage, control-flow distance, and value-flow influence.

\subsubsection*{\textbf{Instrumentation for Coverage Feedback }}
Traditional grey-box fuzzers indiscriminately gather coverage feedback from all basic blocks. In contrast, \tool selectively acquires coverage feedback solely from the sliced basic blocks related to the targets. Specifically, \tool first conducts slicing, using the target locations as the slicing criterion. Subsequently, it instruments only these sliced basic blocks to obtain coverage feedback. For a program $P$ and a set of targets $T$, we define the slicing of $T$ as follows:
\begin{equation}
    \small
    \Psi (P, T) = SC(P, T) \vee SV(P, T)
\end{equation}
where $SC(P, T)$ is the control-flow slicing and $SV(P, T)$ is the value-flow slicing.

Control-flow and value-flow slicing are both accomplished through a breadth-first backward traversal starting from the target location. Initially, \tool performs function-level slicing on the target program using the function call graph and simultaneously extracts the call-site basic blocks invoking the sliced functions. Subsequently, for each sliced function, \tool conducts control-flow slicing. It uses the call-site basic blocks in the control-flow graph as the slicing criterion. Regarding value-flow slicing, \tool begins the operation on the value-flow graph, starting from the nodes where the target data is located.

\subsubsection*{\textbf{Instrumentation for Distance Feedback}} \label{def:sbb}
In \tool, only boundary basic blocks are instrumented to provide  distance feedback. 
Given a program $P$ and all the sliced basic blocks $\Psi(P)$ of the target locations, the set of potential \textbf{Boundary Basic Blocks} of $P$, denoted as $\Omega(P)$), is defined as follows:
\begin{equation}
\small
\begin{split}
    \Omega(P) = \{& bb\in \Psi(P) \mid (Successors(bb) = NULL) \lor \\
                  &(\exists sbb\in Successors(bb) : sbb\notin \Psi (P)) \}
\end{split}
\end{equation}
where $Successors(bb)$ refers to the set of basic blocks that are the immediate successors of the basic block $bb$. 

This definition embodies two crucial aspects. First, a boundary basic block must be among the sliced basic blocks. This ensures it contains target-relevant code and can provide valid distance feedback. Second, a boundary basic block either lacks successor nodes or has only non-sliced successor nodes, signifying its role as the final block in the execution path for distance feedback.

Leveraging boundary basic blocks, \tool can acquire critical distance information, whether the program execution is about to return to the sliced region or deviate from it. Instrumenting non-boundary basic blocks on target-reachable paths not only fails to yield extra valuable distance feedback but also imposes unnecessary runtime overhead. This insight aligns with the results of WindRanger (\cite{du_windranger_2022}) and Selectfuzz (\cite{luo_selectfuzz_2023}), both of which highlight that only deviation basic blocks are essential for analysis.

\subsubsection*{\textbf{Instrumentation for Value-flow Influence Feedback }}
\tool instruments all basic blocks with a value-flow influence score, driven by two key considerations. First, since the impact of each basic block along the execution path on the target data must be factored in, the value-flow influence score of a seed is computed by aggregating the scores of all relevant basic blocks on that path. Second, following \tool's thorough static analysis, only a limited number of basic blocks possess a value-flow influence score. Instrumenting these blocks will not cause a substantial decline in runtime performance.

% 实现===================================================================
\section{Implementation}
\label{sec:implementation}
We developed a prototype of \tool based on AFLGo \cite{bohme_directed_2017}, LLVM \cite{lattner_llvm_2004}, and SVF \cite{SVF}. Specifically, we constructed \tool's static analyzer with LLVM and SVF, writing approximately 1,300 lines of C++ code. This implementation encompasses hybrid slicing, control-flow distance calculation, value-flow influence score calculation, and the identification of boundary basic blocks.
For hybrid feedback acquisition, we added 480 lines of C++ code to AFLGo's LLVM pass for selective instrumentation. Additionally, we modified AFLGo's scheduling algorithm by adding 460 lines and deleting 130 lines of C code. In total, \tool consists of 10,832 lines of C and C++ code.

The source code for our prototype is available at \url{https://github.com/HF-DGF/HF-DGF}.
% https://github.com/HF-DGF/HF-DGF 

% 行号统计
% config.h: +32  -3
% afl-fuzz.c:  +236  -126
% llvm_mode/afl-llvm-rt.o.c: +192
% llvm_mode/afl-llvm-pass.so.cc: +430  -76
% llvm_mode/afl-clang-fast.c: +49   -38
% slicer: 1356
% 8558 + 547 + 371 + 1356 = 

% 实验评估
\section{Evaluation} \label{sec:evaluation}
In this section, we assess the performance of \tool using real-world vulnerabilities and seek answers to the following research questions:
\begin{itemize}
\item \textbf{RQ1}: How effective is \tool at triggering known vulnerabilities?
\item \textbf{RQ2}: How accurately can \tool direct the fuzzing in terms of code coverage?
\item \textbf{RQ3}: How efficient is the static analysis performed by \tool?
\item \textbf{RQ4}: How effective is the selective instrumentation strategy adopted by \tool?
\item \textbf{RQ5}: What impacts do each of the components have on the overall performance of \tool?
\end{itemize}

% 评估实验设置
\subsection{Evaluation Setup}
\subsubsection*{\textbf{Baselines}}
We compare \tool with the following four state-of-the-art fuzzers, which are publicly available:
\begin{itemize}
\item AFL \cite{AFL}: v2.57b, Jul 5, 2020.
\item AFLGo \cite{bohme_directed_2017}: commit b170fad, May 8, 2022.
\item WindRanger \cite{du_windranger_2022}:  Docker hash 8614ceb.
\item DAFL \cite{kim_dafl_2023}: commit bdfed39, Dec 28, 2023.
\item Beacon \cite{huang_beacon_2022}: Docker hash a09c8cb.
\end{itemize}

We excluded Hawkeye \cite{chen_hawkeye_2018} and Selectfuzz \cite{luo_selectfuzz_2023} from our comparison for the following reasons. Hawkeye is not publicly available. Regarding Selectfuzz, despite being partially open-source, we were unable to obtain and compile its static analyzer, which contains the core logic.
We also attempt to evaluate the performance of AFL++ \cite{aflplusplus}. However, probably due to certain configuration-related factors, only the CVE-2017-5969 vulnerability was triggered, while other evaluation targets remained untriggered. Consequently, AFL++ is not included in Table \ref{tab:tte_mix}.

\subsubsection*{\textbf{Benchmark}}
\begin{table}[ht]
  \centering
  \small
  \renewcommand{\arraystretch}{1.09}
  \tabcolsep=2pt
  \caption{Vulnerability Information. The "Type" column indicates the vulnerability types using these abbreviations: \textbf{BOF} (Buffer Overflow), \textbf{ND} (Null Dereference), \textbf{SO} (Stack Overflow), \textbf{UAF} (Use-After-Free), \textbf{IO} (Integer Overflow), \textbf{OOM} (Out of Memory).}

  \begin{tabular}{llllcr}
    \toprule
    \textbf{Project} & \textbf{Program} & \textbf{Version} & \textbf{CVE} & \textbf{Target} & \textbf{Type} \\
    \midrule
    \multirow{17}[4]{*}{LibMing} & \multirow{17}[4]{*}{swftophp} & \multirow{7}[2]{*}{0.4.7} & 2016-9827 & outputscript.c:1687 & BOF \\
          &       &       & 2016-9829 & parser.c:1656 & ND \\
          &       &       & 2016-9831 & parser.c:66 & BOF \\
          &       &       & 2017-9988 & parser.c:2995 & ND \\
          &       &       & 2017-11728 & decompile.c:1699 & ND \\
          &       &       & 2017-11729 & decompile.c:1440 & ND \\
          &       &       & 2017-7578 & parser.c:68 & BOF \\
\cmidrule{3-6}          &       & \multirow{10}[2]{*}{0.4.8} & 2018-7868 & decompile.c:398 & UAF \\
          &       &       & 2018-8807 & decompile.c:349 & UAF \\
          &       &       & 2018-8962 & decompile.c:398 & UAF \\
          &       &       & 2018-11095 & decompile.c:1843 & UAF \\
          &       &       & 2018-11225 & decompile.c:2015 & BOF \\
          &       &       & 2018-11226 & decompile.c:2015 & BOF \\
          &       &       & 2018-20427 & decompile.c:425 & ND \\
          &       &       & 2019-9114 & decompile.c:259 & BOF \\
          &       &       & 2019-12982 & decompile.c:3120 & BOF \\
          &       &       & 2020-6628 & decompile.c:2015 & BOF \\
    \midrule
    \multirow{2}[2]{*}{Lrzip} & \multirow{2}[2]{*}{lrzip} & \multirow{2}[2]{*}{0.631} & 2017-8846 & stream.c:1747 & UAF \\
          &       &       & 2018-11496 & stream.c:1756 & UAF \\
    \midrule
    \multirow{17}[14]{*}{Bintuils} & \multirow{6}[2]{*}{cxxfilt} & \multirow{6}[2]{*}{2.6} & 2016-4487 & cplus-dem.c:4319 & ND \\
          &       &       & 2016-4489 & cplus-dem.c:3007 & UAF \\
          &       &       & 2016-4490 & cp-demangle.c:1596 & IO \\
          &       &       & 2016-4491 & cp-demangle.c:5472 & SO \\
          &       &       & 2016-4492 & cplus-dem.c:3606 & IO \\
          &       &       & 2016-6131 & cplus-dem.c:4231 & SO \\
\cmidrule{2-6}          & \multirow{3}[2]{*}{objcopy} & \multirow{3}[2]{*}{2.8} & 2017-8393 & elf.c:3562 & BOF \\
          &       &       & 2017-8394 & objcopy.c:1482 & ND \\
          &       &       & 2017-8395 & cache.c:336 & ND \\
\cmidrule{2-6}          & \multirow{5}[4]{*}{objdump} & \multirow{4}[2]{*}{2.8} & 2017-8392 & libbfd.c:548 & ND \\
          &       &       & 2017-8396 & libbfd.c:615 & BOF \\
          &       &       & 2017-8397 & reloc.c:885 & BOF \\
          &       &       & 2017-8398 & dwarf.c:484 & BOF \\
\cmidrule{3-6}          &       & 2.31.1 & 2018-17360 & libbfd.c:656 & BOF \\
\cmidrule{2-6}          & strip & 2.7   & 2017-7303 & elf.c:1252 & ND \\
\cmidrule{2-6}          & nm    & 2.9   & 2017-14940 & elf.c:8205 & OOM \\
\cmidrule{2-6}          & readelf & 2.9   & 2017-16828 & dwarf.c:7019 & IO \\
    \midrule
    \multirow{3}[2]{*}{Libxml2} & \multirow{3}[2]{*}{xmllint} & \multirow{3}[2]{*}{2.9.4} & 2017-5969 & valid.c:1181 & ND \\
          &       &       & 2017-9047 & valid.c:1279 & BOF \\
          &       &       & 2017-9048 & valid.c:1323 & BOF \\
    \midrule
    \multirow{2}[2]{*}{Libjpeg} & \multirow{2}[2]{*}{cjpeg} & 1.5.90 & 2018-14498 & rdbmp.c:209 & BOF \\
          &       & 2.0.4 & 2020-13790 & rdppm.c:434 & BOF \\
    \midrule
    \textbf{\# Total} & \textbf{10} &       & \textbf{41} &       & \textbf{6} \\
    \bottomrule
    \end{tabular}
  \label{tab:vuls}
\end{table}

We evaluated fuzzers' performance using various security vulnerabilities in C programs. Specifically, we adopted the 41 vulnerabilities from DAFL \cite{kim_dafl_2023}, detailed in Table \ref{tab:vuls}.
As is common in vulnerability-exposing fuzzing efforts, we built all target binaries with ASAN (AddressSanitizer). But since Beacon doesn't support fuzzing ASAN-compiled target binaries, we ran an extra experiment without ASAN to compare with Beacon.

\subsubsection*{\textbf{Environment}}
The experiments were conducted on machines equipped with an AMD(R) EPYC 9354 CPU @ 3.80GHz, 256GB of memory, and running Ubuntu 20.04 LTS. Each fuzzing session took place within a Docker container \cite{docker_what_nodate}, with 1 CPU core and 4GB of memory allocated. Of the 64 available CPU cores, only 54 were used, all within the same fuzzing session.
For all fuzzers, the fuzzing phase was given a 24-hour time budget. To reduce the impact of fuzzing's inherent randomness, the experiments were repeated 20 times, and the median value was computed.

\subsection{Time-to-Exposure (RQ1)}\label{exp:TTE}

\begin{table*}[!ht]
  \centering
  % \footnotesize
  \small
  \renewcommand{\arraystretch}{1.15}
  \tabcolsep=3.2pt
  \caption{Crash reproduction results of \tool and the baseline tools.
  Each TTE represents the median value from 20 repeated experiments. \textbf{N.A.} indicates that the tool was unable to produce a median TTE, meaning it failed to reproduce the bug in more than half of the repeated experiments. The best result for each target (row) is highlighted in bold font. If no tool can successfully obtain the median TTE for a target, it is marked with bold font and an asterisk in the CVE column. \textbf{\#TTE} denotes the number of successfully produced median TTEs by the tool. \textbf{\#Best Perf.} denotes the number of best performances achieved by the tool among all fuzzers.}
    \begin{tabular}{lrrrrrrrrr|rrrrr}
    \toprule

    \multicolumn{10}{c|}{With ASAN}    & \multicolumn{5}{c}{Without ASAN} \\
    \cmidrule{2-15}       & \multicolumn{2}{c}{\textbf{AFL}} & \multicolumn{2}{c}{\textbf{AFLGo}} & \multicolumn{2}{c}{\textbf{WindRanger}} & \multicolumn{2}{c}{\textbf{DAFL}} & \textbf{\Toolname} & \multicolumn{2}{c}{\textbf{DAFL}} & \multicolumn{2}{c}{\textbf{Beacon}} & \textbf{\Toolname} \\

    % 中间有间隔的小横线
    \cmidrule(r{4pt}){2-3}     \cmidrule(l{4pt}r{4pt}){4-5}
    \cmidrule(l{4pt}r{4pt}){6-7}     \cmidrule(l{4pt}r{4pt}){8-9}  \cmidrule(l{4pt}r{2pt}){10-10}  \cmidrule(l{2pt}r{4pt}){11-12}  \cmidrule(l{4pt}r{4pt}){13-14}  \cmidrule(l{4pt}){15-15}  

    \textbf{CVE} & \textbf{TTE} & \textbf{Factor} & \textbf{TTE} & \textbf{Factor} & \textbf{TTE} & \textbf{Factor} & \textbf{TTE} & \textbf{Factor} & \textbf{TTE} & \textbf{TTE} & \textbf{Factor} & \textbf{TTE} & \textbf{Factor} & \textbf{TTE} \\

    \midrule
    2016-9827 & 88 & 3.52  & 68 & 2.72  & 59 & 2.36  & 52 & 2.08  & \textbf{25} & 130 & 1.29  & 2110 & 20.89  & \textbf{101} \\
    2016-9829 & 444 & 5.48  & 137 & 1.69  & 1865 & 23.02  & 101 & 1.25  & \textbf{81} & \textbf{1015} & -  & 11335 & -  & N.A. \\
    2016-9831 & 280 & 3.73  & 267 & 3.56  & N.A. & -  & 137 & 1.83  & \textbf{75} & 335 & 3.10  & 223 & 2.06  & \textbf{108} \\
    2017-9988 & \textbf{157} & 0.23  & 3455 & 5.10  & 1275 & 1.88  & 1128 & 1.66  & 678 & 6480 & 8.47  & 2892 & 3.78  & 765 \\
    2017-11728 & 1360 & 10.97  & 2578 & 20.79  & 2001 & 16.14  & 202 & 1.63  & \textbf{124} & N.A. & -  & N.A. & -  & \textbf{537} \\
    2017-11729 & 518 & 9.42  & 413 & 7.51  & 714 & 12.98  & 56 & 1.02  & \textbf{55} & 280 & 2.57  & 3430 & 31.47  & \textbf{109} \\
    2017-7578 & 517 & 8.48  & 185 & 3.03  & N.A. & -  & \textbf{37} & 0.61  & 61 & 1546 & 7.47  & 1366 & 6.60  & \textbf{207} \\
    2018-7868 & N.A. & -  & N.A. & -  & N.A. & -  & 11746 & 2.19  & \textbf{5359} & N.A. & -  & N.A. & -  & \textbf{6273} \\
    \textbf{2018-8807*} & N.A. & -  & N.A. & -  & N.A. & -  & N.A. & -  & N.A. & N.A. & -  & N.A. & -  & N.A. \\
    \textbf{2018-8962*} & N.A. & -  & N.A. & -  & N.A. & -  & N.A. & -  & N.A. & N.A. & -  & N.A. & -  & N.A. \\
    2018-11095 & 2799 & 7.46  & 2523 & 6.73  & 837 & 2.23  & 537 & 1.43  & \textbf{375} & 2088 & 1.91  & 6065 & 5.54  & \textbf{1094} \\
    2018-11225 & N.A. & -  & 16849 & 1.75  & N.A. & -  & N.A. & -  & \textbf{9626} & N.A. & -  & 63641 & 17.19  & \textbf{3702} \\
    2018-11226 & \textbf{4313} & 0.61  & N.A. & -  & N.A. & -  & 5928 & 0.83  & 7128 & N.A. & -  & N.A. & -  & \textbf{5015} \\
    2018-20427 & 2761 & 1.49  & 4602 & 2.49  & 4946 & 2.67  & 9701 & 5.24  & \textbf{1851} & 29003 & 25.76  & 1574 & 1.40  & \textbf{1126} \\
    2019-9114 & 19310 & -  & \textbf{7962} & -  & 84885 & -  & N.A. & -  & N.A. & N.A. & -  & 16681 & 2.28  & \textbf{7308} \\
    2019-12982 & N.A. & -  & 13695 & 2.42  & N.A. & -  & N.A. & -  & \textbf{5668} & N.A. & -  & N.A. & -  & N.A. \\
    2020-6628 & 72988 & 4.02  & 31732 & 1.75  & N.A. & -  & 41132 & 2.26  & \textbf{18167} & 25726 & 4.25  & N.A. & -  & \textbf{6047} \\
    \textbf{2017-8846*} & N.A. & -  & N.A. & -  & N.A. & -  & N.A. & -  & N.A. & N.A. & -  & N.A. & -  & N.A. \\
    2018-11496 & 17 & 5.67  & 26 & 8.67  & 9  & 3.00  & 8  & 2.67  & \textbf{3} & \textbf{2} & 1.00  & N.A. & -  & \textbf{2} \\
    2016-4487 & 407 & 5.02  & 259 & 3.20  & 183 & 2.26  & 143 & 1.77  & \textbf{81} & 134 & 3.62  & 101 & 2.73  & \textbf{37} \\
    2016-4489 & 533 & 3.68  & 826 & 5.70  & 231 & 1.59  & 1137 & 7.84  & \textbf{145} & 1451 & 38.18  & 191 & 5.03  & \textbf{38} \\
    2016-4490 & 178 & 2.54  & 445 & 6.36  & 99 & 1.41  & \textbf{46} & 0.66  & 70 & 39 & 3.25  & 42 & 3.50  & \textbf{12} \\
    2016-4491 & N.A. & -  & N.A. & -  & N.A. & -  & N.A. & -  & N.A. & N.A. & -  & N.A. & -  & \textbf{11710} \\
    2016-4492 & 1909 & 1.12  & 3385 & 1.98  & 1003 & 0.59  & \textbf{891} & 0.52  & 1708 & 817 & 1.56  & 753 & 1.43  & \textbf{525} \\
    2016-6131 & N.A. & -  & N.A. & -  & N.A. & -  & N.A. & -  & \textbf{50063} & N.A. & -  & N.A. & -  & N.A. \\
    2017-8393 & \textbf{707} & 0.04  & 896 & 0.05  & 1280 & 0.07  & 1228 & 0.07  & 17601 & N.A. & -  & N.A. & -  & N.A. \\
    2017-8394 & 643 & 2.82  & 668 & 2.93  & 946 & 4.15  & 251 & 1.10  & \textbf{228} & N.A. & -  & 1458 & 19.97  & \textbf{73} \\
    2017-8395 & 88 & 0.02  & 86 & 0.02  & 101 & 0.03  & \textbf{8} & 0.00  & 3879 & N.A. & -  & \textbf{12} & 0.00  & 3944 \\
    2017-8392 & 298 & 0.09  & 341 & 0.10  & N.A. & -  & \textbf{15} & 0.00  & 3270 & N.A. & -  & 48339 & 1.08  & \textbf{44661} \\
    \textbf{2017-8396*} & N.A. & -  & N.A. & -  & N.A. & -  & N.A. & -  & N.A. & N.A. & -  & N.A. & -  & N.A. \\
    \textbf{2017-8397*} & N.A. & -  & N.A. & -  & N.A. & -  & N.A. & -  & N.A. & N.A. & -  & N.A. & -  & N.A. \\
    2017-8398 & \textbf{709} & 0.04  & 3551 & 0.22  & 1138 & 0.07  & 4359 & 0.27  & 16384 & N.A. & -  & N.A. & -  & N.A. \\
    \textbf{2018-17360*} & N.A. & -  & N.A. & -  & N.A. & -  & N.A. & -  & N.A. & N.A. & -  & N.A. & -  & N.A. \\
    2017-7303 & 91 & 0.02  & \textbf{84} & 0.02  & 153 & 0.03  & 166 & 0.03  & 4934 & 221 & 0.26  & \textbf{14} & 0.02  & 840 \\
    \textbf{2017-14940*} & N.A. & -  & N.A. & -  & N.A. & -  & N.A. & -  & N.A. & N.A. & -  & N.A. & -  & N.A. \\
    2017-16828 & 145 & 4.68  & 124 & 4.00  & 212 & 6.84  & 131 & 4.23  & \textbf{31} & N.A. & -  & N.A. & -  & N.A. \\
    2017-5969 & 198 & 33.00  & 192 & 32.00  & 1292 & 215.33  & 96 & 16.00  & \textbf{6} & 348 & 174.00  & 51 & 25.50  & \textbf{2} \\
    2017-9047 & N.A. & -  & N.A. & -  & N.A. & -  & \textbf{16067} & 0.30  & 52814 & N.A. & -  & \textbf{4340} & 0.16  & 27025 \\
    2017-9048 & N.A. & -  & N.A. & -  & N.A. & -  & \textbf{5214} & -  & N.A. & N.A. & -  & N.A. & -  & N.A. \\
    2018-14498 & N.A. & -  & N.A. & -  & N.A. & -  & \textbf{7382} & -  & N.A. & N.A. & -  & N.A. & -  & N.A. \\
    \textbf{2020-13790*} & N.A. & -  & N.A. & -  & N.A. & -  & N.A. & -  & N.A. & N.A. & -  & N.A. & -  & N.A. \\
    % \midrule
    \cmidrule(r{4pt}){1-1} \cmidrule(l{4pt}r{4pt}){2-3}     \cmidrule(l{4pt}r{4pt}){4-5}
    \cmidrule(l{4pt}r{4pt}){6-7}     \cmidrule(l{4pt}r{4pt}){8-9}  \cmidrule(l{4pt}r{2pt}){10-10}  \cmidrule(l{2pt}r{4pt}){11-12}  \cmidrule(l{4pt}r{4pt}){13-14}  \cmidrule(l{4pt}){15-15}  
    
    \textbf{\# TTE (33/41)} & \multicolumn{2}{c}{\textbf{25}} & \multicolumn{2}{c}{\textbf{26}} & \multicolumn{2}{c}{\textbf{20}} & \multicolumn{2}{c}{\textbf{28}} & \textbf{29} & \multicolumn{2}{c}{\textbf{16}} & \multicolumn{2}{c}{\textbf{20}} & \textbf{25} \\
    \textbf{\# Best Perf.} & \multicolumn{2}{c}{\textbf{4}} & \multicolumn{2}{c}{\textbf{2}} & \multicolumn{2}{c}{\textbf{0}} & \multicolumn{2}{c}{\textbf{8}} & \textbf{18} & \multicolumn{2}{c}{\textbf{2}} & \multicolumn{2}{c}{\textbf{3}} & \textbf{21} \\
    \bottomrule
    \end{tabular}%
  \label{tab:tte_mix}%
\end{table*}%

We initiate the evaluation of \tool's effectiveness in terms of Time-To-Exposure (TTE) through the crash reproduction experiment. The results are presented in Table~\ref{tab:tte_mix}. When a fuzzer fails to reproduce the target crash within 24 hours in over half of the 20 trials, making it impossible to obtain the median data, we mark the result as Not Available (N.A.).

After excluding the 9 cases where no fuzzing tool could obtain a median value, \tool was able to report a median Time-To-Exposure (TTE) for 29 benchmarks. By comparison, AFL, AFLGo, WindRanger, and DAFL reported median TTEs for 25, 26, 20, and 28 benchmarks respectively. Additionally, among the remaining 32 cases, \tool outperformed the other fuzzing tools in 18 instances. In contrast, AFL, AFLGo, WindRanger, and DAFL outperformed the rest in only 4, 2, 0, and 8 cases respectively. For the 20 benchmarks where all the tools successfully reported a median TTE, on average, \tool was roughly 5.05 times faster than AFL, 5.79 times faster than AFLGo, a remarkable 73.75 times faster than WindRanger, and 2.56 times faster than DAFL.
Furthermore, for the target vulnerabilities CVE-2018-7868, CVE-2018-11225, and CVE-2019-12982, where all other fuzzing tools faced challenges in reproducing the crashes, \tool demonstrated a significant performance advantage over them.

In the supplementary experiment without AddressSanitizer, \tool reported TTEs for 25 benchmarks. In comparison, DAFL and Beacon did so for only 16 and 20 benchmarks, respectively. Additionally, \tool outperformed DAFL and Beacon in 21 cases, while DAFL and Beacon were superior in just 2 and 3 cases, respectively. In scenarios where all three tools provided median TTEs, \tool was on average 20.87 times faster than DAFL and 8.45 times faster than Beacon.

% 1review2: 这里增加按照漏洞类型进行统计结果，分析模糊测试器对漏洞亲和度
We further investigated the fuzzers' affinity for different types of vulnerabilities. The results are shown in Table \ref{tab:vul_type}. \tool surpasses all others in reproducing four cases of use-after-free (UAF) vulnerabilities, while no other fuzzer can reproduce the remaining three UAF vulnerabilities. Only \tool and DAFL successfully reproduce one of the two Stack Overflow vulnerabilities, and all the tested fuzzers fail to reproduce the other. Additionally, none of the evaluated fuzzers show the ability to reproduce the Out-of-Memory vulnerability. For various vulnerability types, \tool exhibits an affinity comparable to that of other tools.

In summary, experimental results show that our proposed \tool markedly boosts the performance of crash reproduction. Leveraging both distance and value-flow influence feedback, \tool can comprehensively explore code segments distant from the target yet affecting its state space. This essentially strengthens \tool's potential to unearth deeply hidden vulnerabilities.

% % Table generated by Excel2LaTeX from sheet 'vul_type_num'
% \begin{table}[htbp]
%   \centering
%   \small
%   \caption{Results of the number of vulnerabilities grouped by type.}
%     \begin{tabular}{lccc}
%     \toprule
%     \textbf{Vulnerability Type} & \textbf{\# Total} & \textbf{\# TTE} & \textbf{\# Best} \\
%     \midrule
%     Buffer Overflow & 17    & 10    & 5 \\
%     Null Dereference & 11    & 11    & 7 \\
%     Use-After-Free & 7     & 4     & 4 \\
%     Integer Overflow & 3     & 3     & 1 \\
%     Stack Overflow & 2     & 1     & 1 \\
%     Out of Memory & 1     & 0     & 0 \\
%     \bottomrule
%     \end{tabular}%
%   \label{tab:vul_type}%
% \end{table}%

% Table generated by Excel2LaTeX from sheet 'vul_type_num (2)'
\begin{table*}[!t]
  \centering
  \small
  \renewcommand{\arraystretch}{1.15}
  \tabcolsep=8pt
  \caption{Crash reproduction results grouped by type. Results are presented in the format \textbf{\#Best/\#TTE}, where \textbf{\#TTE} represents the number of median TTE successfully produced by the tool, and \textbf{\#Best} represents the number of best performences achieved by the tool among all the fuzzers.}
    \begin{tabular}{lc|ccccc|ccc}
    \toprule
    \multicolumn{1}{c}{\multirow{2}[4]{*}{\textbf{Vulnerability Type}}} & \multirow{2}[4]{*}{\textbf{\# Total}} & 
    \multicolumn{5}{c|}{With ASAN}    & \multicolumn{3}{c}{Without ASAN} \\
\cmidrule{3-10}          &       & \textbf{AFL} & \textbf{AFLGo} & \textbf{WindRanger} & \textbf{DAFL} & \textbf{\tool} & \textbf{DAFL} & \textbf{Beacon} & \textbf{\tool} \\

    \midrule
    Buffer Overflow & 17    & 3 / 8 & 1 / 9 & 0 / 4 & 4 / 10 & 5 / 10 & 0 / 4 & 1 / 6 & 7 / 8 \\
    Null Dereference & 11    & 1 / 11 & 1 / 11 & 0 / 10 & 2 /11 & 7 / 11 & 1 / 7 & 2 / 10 & 7 / 10 \\
    Use-After-Free & 7     & 0 / 3 & 0 / 3 & 0 / 3 & 0 / 4 & 4 / 4 & 1 / 3 & 0 / 2 & 4 / 4 \\
    Integer Overflow & 3     & 0 / 3 & 0 / 3 & 0 / 3 & 2 / 3 & 1 / 3 & 0 / 2 & 0 / 2 & 2 / 2 \\
    Stack Overflow & 2     & 0 / 0 & 0 / 0 & 0 / 0 & 1 / 1 & 1 / 1 & 0 / 0 & 0 / 0 & 1 / 1 \\
    Out of Memory & 1     & 0 / 0 & 0 / 0 & 0 / 0 & 0 / 0 & 0 / 0 & 0 / 0 & 0 / 0 & 0 / 0 \\
    \bottomrule
    \end{tabular}%
  \label{tab:vul_type}%
\end{table*}%

% 覆盖率，定向性
\subsection{Code Coverage (RQ2)}
\label{eva:coverage}
\begin{figure*}[htpb]
\centering
\includegraphics[width=2\columnwidth]{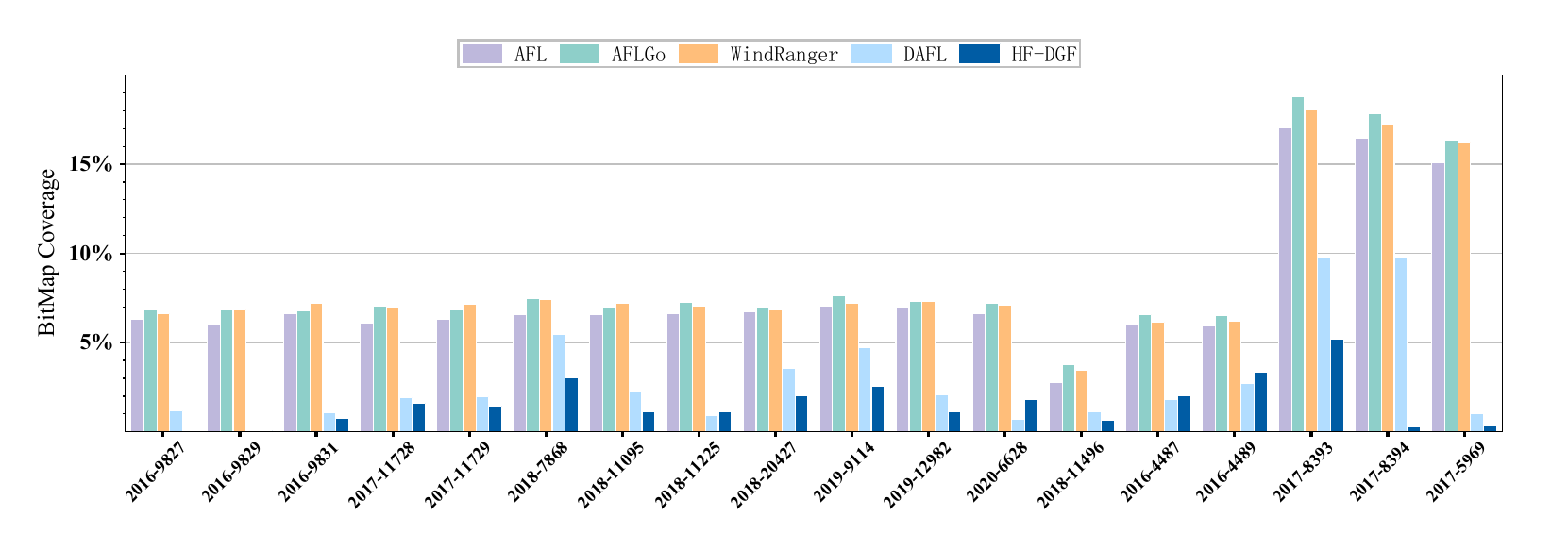}
\caption{BitMap Coverage}
\label{fig:cov}
\end{figure*}

Subsequently, we evaluated \tool's directional ability in fuzzing. From the 41 evaluation cases mentioned above, we chose the 18 top-performing targets according to the TTE \tool achieved. Next, we acquired \tool's coverage metrics and compared them with those of all baseline fuzzers. The representative average, based on 20 iterations, is shown in Fig.~\ref{fig:cov}.

In all 18 test cases, \tool exhibited the lowest code coverage, being 16.81 times lower than AFL, 18.45 times lower than AFLGo, 18.16 times lower than WindRanger, and 3.71 times lower than DAFL. Nevertheless, \tool triggered the target vulnerabilities in all cases. This shows that even with a narrow exploration scope, \tool did not overlook any of these vulnerabilities.

The proposed static analysis method significantly contributes to this success. By precisely computing control-flow distances and value-flow influence scores, \tool effectively prioritizes fuzzing time towards reaching the target location and exploring the target state space. It focuses on parts closely linked to the target vulnerability and reduces the exploration of irrelevant code.

% 静态分析性能分析
\subsection{Performance of Static Analyzer (RQ3)}

In directed grey-box fuzzing, the time for static analysis is usually not a one-off cost, especially in continuous fuzzing with frequent code modifications. Static analysis must be repeated after each modification. Thus, we evaluate both the time overhead of static analysis and the time proportion of each component in it.

% 静态分析时间对比
\subsubsection*{\textbf{Overhead of Static Analyzer}}
\begin{figure}[htpb]
\centering
\includegraphics[width=\columnwidth]{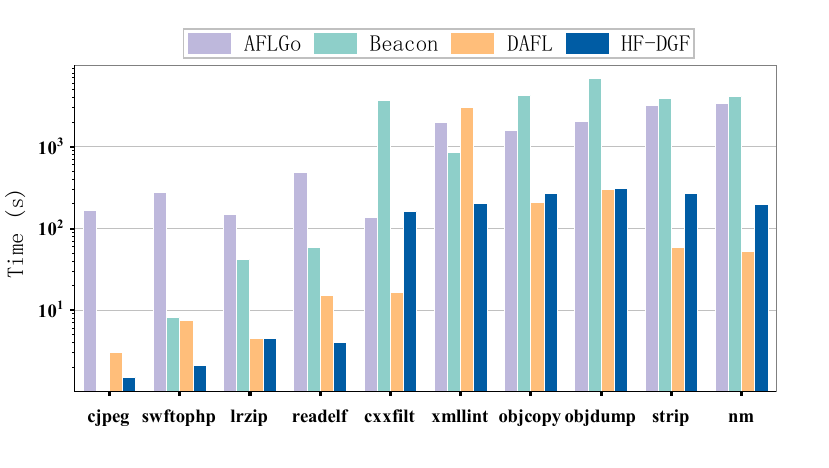}
\caption{Static Analyzer Time}
\label{fig:sa_time}
\end{figure}

Figure \ref{fig:sa_time} compares the static-analysis-phase times of AFLGo, Beacon, DAFL, and our proposed \tool. Evidently, \tool is far more efficient in static analysis, needing much less time than the others. On average, \tool analyzes 44.56 times faster than AFLGo, 12.89 times faster than Beacon, and 2.75 times faster than DAFL. Its quicker static analysis allows the fuzzing phase to start earlier, which is crucial for urgent security testing.

% 静态分析各组件时间占比
\subsubsection*{\textbf{Overhead of Each Component}}
\begin{figure}[htpb]
\centering
\includegraphics[width=1.02\columnwidth]{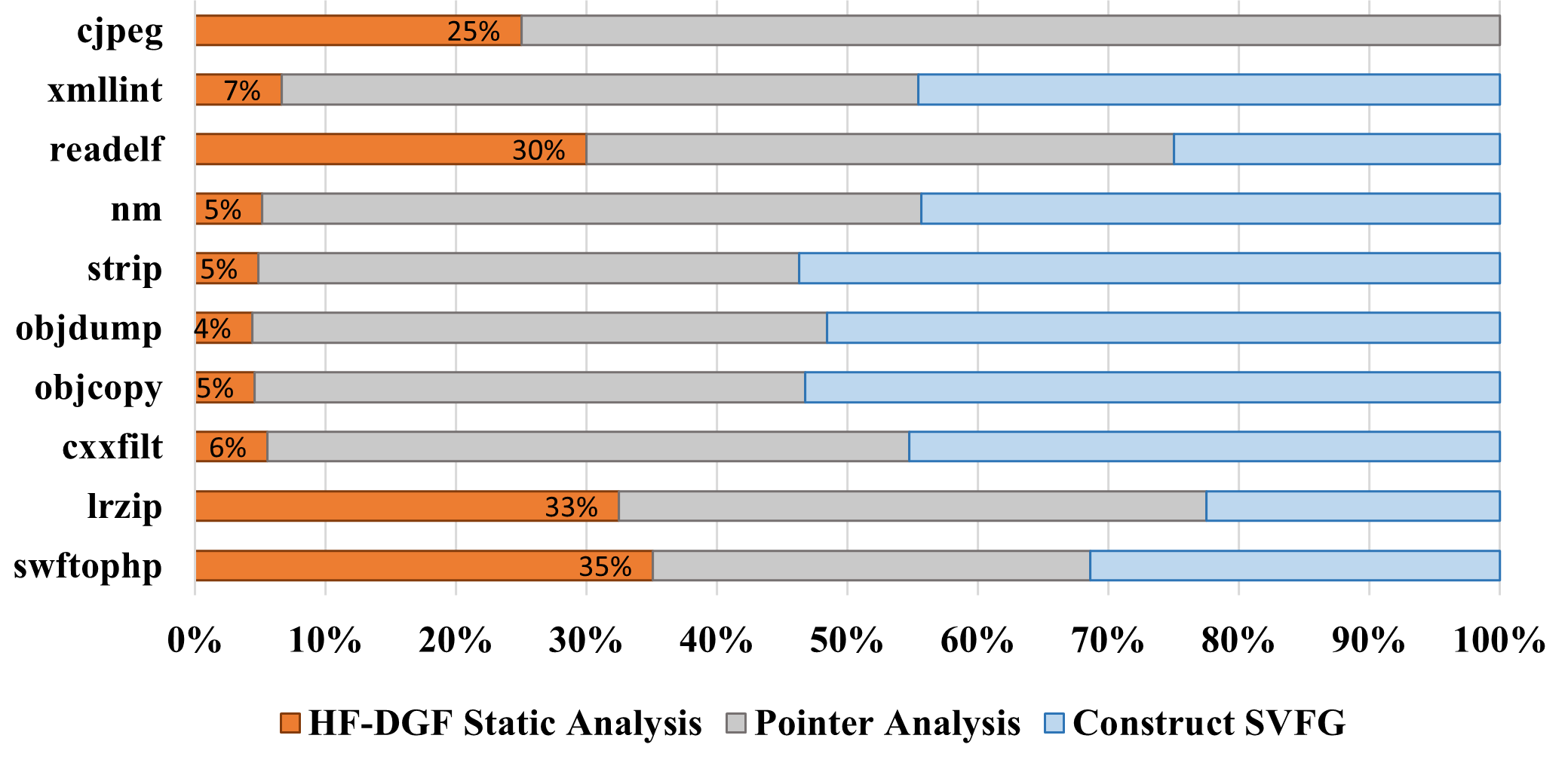}
\caption{Static Analysis Component Time Consumption}
\label{fig:sa_com}
\end{figure}

We evaluated the time consumption of each component in \tool's static analysis phase, as shown in Fig.~\ref{fig:sa_com}. Pointer analysis and constructing the Sparse Value-Flow Graph (SVFG) took up most of the time, accounting for 45\% and 35\% respectively. These substantial time costs highlight potential areas for optimizing static analysis. By improving the algorithms for pointer analysis and SVFG construction, we anticipate further cutting down the static-analysis time and boosting the overall efficiency of the fuzzing process.

% 选择性插桩策略的有效性
\subsection{Effectiveness of Selective Instrumentation Strategy (RQ4)}
We evaluated the effectiveness of \tool's selective instrumentation strategy on performance improvement, and the relevant statistical results are presented in Table~\ref{tab:inst_mini}.

For each target vulnerability, we quantified several key metrics: the total number of basic blocks in the target program (\#BB), the number of basic blocks instrumented for coverage feedback (\#Cov-Ins), the number of target-reachable basic blocks (\#Dis-BB), the number of boundary basic blocks instrumented for control-flow distance feedback (\#Dis-Ins), and the number of basic blocks instrumented for value-flow feedback (\#Val-Ins).

\begin{table}[t]
  \centering
  \small
  \renewcommand{\arraystretch}{1.15}
  \tabcolsep=6pt
  \caption{Impact of Selective Instrumentation Strategy. \textbf{\#BB} denotes the total quantity of basic blocks, \textbf{\#Cov-Ins} represents the number of basic blocks instrumented for coverage feedback, \textbf{\#Dis-BB} refers to the number of target-reachable basic blocks, \textbf{\#Dis-Ins} is the number of boundary basic blocks instrumented for control-flow distance feedback, and \textbf{\#Val-Ins} stands for the number of basic blocks instrumented to offer value-flow feedback.}
    \begin{tabular}{lrrrrr}
    \toprule
    \textbf{Program} & \textbf{\#BB} &\textbf{\#Cov-Ins} &\textbf{\#Dis-BB} &\textbf{\#Dis-Ins} & \textbf{\#Val-Ins} \\
    \midrule
    swftophp & 3700  & 450   & 379   & 87    & 96  \\
    lrzip & 9198  & 772   & 337   & 42    & 31  \\
    cxxfilt & 42170  & 950   & 750   & 180   & 123  \\
    objcopy & 51491  & 10633  & 10404  & 2397  & 350  \\
    objdump & 58730  & 15979  & 15730  & 3648  & 126  \\
    strip & 50788  & 15434  & 15212  & 3529  & 36  \\
    nm    & 45190  & 14786  & 14771  & 3326  & 18  \\
    readelf & 18850  & 533   & 533   & 31    & 2  \\
    xmllint & 69907  & 25683  & 25355  & 6317  & 192  \\
    cjpeg & 5888  & 55    & 52    & 15    & 15  \\
    \midrule
    \textbf{Ratio } &       & \textbf{23.96\%}&       & \textbf{23.43\%}& \textbf{0.28\%} \\
    \bottomrule
    \end{tabular}%
  \label{tab:inst_mini}%
\end{table}%

The results show that our selective instrumentation strategy substantially cuts down the number of basic blocks that need to be instrumented for coverage feedback (by 76.04\%) and control-flow distance feedback (by 76.57\%). Notably, even though all basic blocks with value-flow influence scores are instrumented, they merely make up 0.28\% of the total basic blocks.

Employing the selective instrumentation strategy, \tool precisely hones its exploration in sliced basic code blocks closely linked to the target location. Although \tool drastically cuts down the number of instrumented basic code blocks, it fully preserves the capacity to unearth target vulnerabilities, as validated in \S \ref{exp:TTE}.

% 消融实验
\subsection{Ablation Study (RQ5)}
\begin{figure}[t]
\centering
\includegraphics[width=\columnwidth]{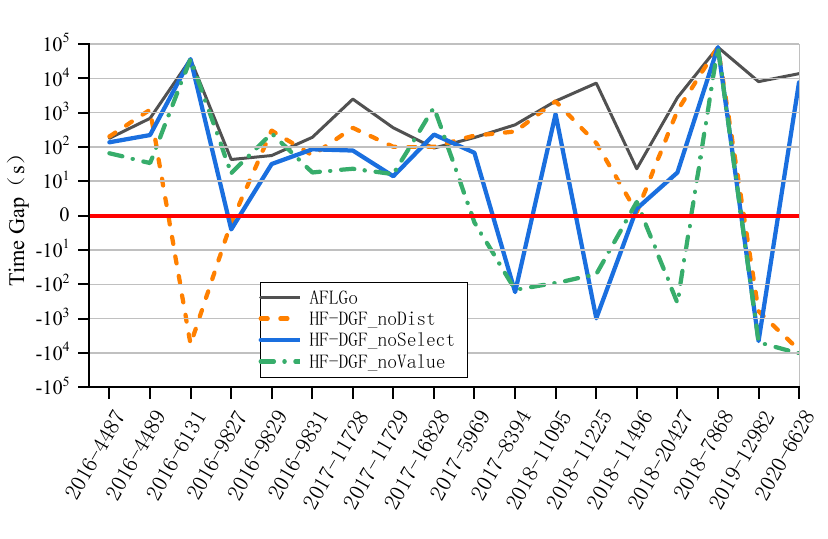}
\caption{Ablation of \Toolname.}
\label{fig:aba}
\end{figure}

In this section, we examine the impact of the techniques employed on \tool's performance. To do so, we instantiated three variants of \tool, as follows:
\begin{itemize}[leftmargin=*]
\item \textbf{\Toolname\_noDist}: This variant omits backward-stepping distance calculation and adopts AFLGo's distance-calculation approach.
\item \textbf{\Toolname\_noSelect}: It lacks the selective instrumentation.
\item \textbf{\Toolname\_noValue}: This one does not incorporate value-flow influence feedback.
\end{itemize}
We compared \tool with its variants on 18 vulnerabilities. The TTE gaps between them and \tool, as shown in Fig.~\ref{fig:aba}, clarify the performance differences. Points above the axis of y = 0 (the solid red line) indicate that a variant takes longer to trigger crashes than \tool; points below it mean a shorter time is taken.

Each technique independently provides positive guidance for \tool, enhancing its performance.
\tool\_noDist, \tool\_noSelect, and \tool\_noValue are 2.08, 3.86, and 7.46 times faster than AFLGo respectively. When combined, these techniques enable \tool to consistently outpace AFLGo, reproducing vulnerabilities 6.98 times faster on average, demonstrating their synergistic effect.

% 讨论==============================================================================================
\section{Discussion}
\label{sec:discussion}
In this section, we discuss the limitations of \tool and possible future work.

\subsection{Solving Complex Constraints}
\tool employs random mutations to generate inputs for solving path constraints and data constraints, but it may not be sufficiently effective in solving complex constraints. Solving complex constraints, which involve reaching deep code locations and inducing specific values for variables, poses a common challenge in fuzzing. Recent studies have proposed various techniques such as symbolic executions \cite{stephens_driller_2016, yun_qsym_2018}, taint tracking \cite{chen_angora_2018, gan_greyone_2020, liang_pata_2022, chen_selectivetaint_2021}, and structure-aware mutations \cite{wang_superion_2019, cao_oddfuzz_2023} to enhance the capability of fuzzers in generating high-quality inputs. Other works trigger sequence-sensitive vulnerabilities by controlling the order of execution. For instance, DDRace \cite{yuan_ddrace_2023} exploits concurrent UAF vulnerability by controlling thread interleaving while SCDF \cite{liang_sequence_2020} and SDFuzz \cite{li_sdfuzz_nodate} generate inputs that can reach the target in order. Integrating \tool with these techniques could enhance its ability to solve complex constraints.

\subsection{The over-tainting issue in static analysis}
Some programs, such as parsers and state machines, connect a large number of functions via dispatch functions. However, typically only one or a few functions are relevant to the target. Existing directed fuzzers often waste a considerable amount of time exploring new but irrelevant functions. \tool try to reduce the misguidance caused by irrelevant functions through hybrid slicing. Though our adopted pointer analysis in slicing is structure-sensitive; however, the current implementation fails to differentiate between different regions of the input, and thus cannot associate functions that parse specific regions accurately. As a result, \tool can still be misled by certain pointers, leading to the inclusion of target-unrelated functions in its exploration scope. For instance, during the evaluation of vulnerabilities like CVE-2017-8395, \tool was found to underperform other baseline fuzzers that do not utilize pointer analysis.

In our future work, we intend to employ advanced static analysis techniques for dispatch function-related pointers, such as roughly determining the reference region through a symbolic execution engine \cite{cadar_klee_2008}. This lightweight approach only requires distinguishing between regions without involving complex solving, thereby avoiding substantial performance overhead.

% 相关工作=========================================================================================================
\section{Related Work}
In this section, we focus on discussing the most related works: control-flow guided directed fuzzing, data-flow guided directed fuzzing, and methods for reducing runtime overhead.

% 基于距离的
\subsection{Control-flow Guided Directed Fuzzing}
DGF enhances directionality by prioritizing and allocating more energy to seeds that are in closer proximity to the target, as measured using control-flow distance. AFLGo \cite{bohme_directed_2017} initially introduced this distance metric and proposed a methodology for computing the distance between a seed trace and target sites. Accurate calculation of the distance is crucial to avoid any bias towards specific traces. Hawkeye \cite{chen_hawkeye_2018} augments the adjacent-function distance by analyzing the frequency and distribution of function calls. WindRanger \cite{du_windranger_2022}, instead of focusing on all basic blocks, calculates the distance through deviation blocks which act as critical basic blocks impeding inputs from reaching the target site, thereby enhancing DGF's efficiency. In contrast to calculating distances during static analysis phase, Parmesan \cite{osterlund_parmesan_2020} computes distances based on dynamically constructed control-flow graphs. PDGF \cite{zhang_predecessor-aware_2023} obtains predecessors through ICFG static analysis and dynamically augments them with execution feedback.

% 数据流增强的
\subsection{Dataflow-Guided Directed Fuzzing}
Data-sensitive vulnerabilities are only triggered under specific program states, not merely by reaching the target location. Consequently, existing work utilizes data flow to guide fuzzing in exploring program states.
According to the stage at which data flow takes effect, existing works can be divided into the following two categories:

\textbf{Static Data Flow Analysis.}
DDFuzz \cite{mantovani_fuzzing_2022} and DataFlow \cite{herrera_dataflow_2023} extend coverage feedback metrics by detecting new data-flow edge coverage. DAFL \cite{kim_dafl_2023} analyzes data dependencies through the Def-Use Graph (DUG) and guides seed scheduling based on semantic relevance scores.
CAFL \cite{CAFL} strengthens DGF by prioritizing crash targets using ordered vulnerability traces from crash dumps and patch analysis. Although its sequential constraint-solving mechanism boosts intentional crash exposure, it restricts path exploration with predefined solution sequences, limiting the discovery of novel vulnerabilities outside established paths.

\textbf{Dynamic Data Flow Tracking.}
GreyOne \cite{gan_greyone_2020} and WindRanger \cite{du_windranger_2022} use probing-based taint analysis to establish associations between input bytes and constraint variables. GreyOne then uses this mapping to decide which bytes to mutate, how to mutate them, and tune the fuzzing evolution direction. WindRanger adopts data-flow-sensitive mutation to better penetrate branch conditions. Nevertheless, runtime taint tracking introduces additional overhead and may not always yield desired outcomes.

\tool integrates the benefits of static analysis and dynamic guidance by computing the value-flow influence of basic blocks during static analysis and accumulating influence at runtime to direct fuzzing, thereby achieving a balance between speed and accuracy.

% 运行时开销降低：路径剪枝+选择性插桩
\subsection{Methods for Reducing Runtime Overhead}
Current research efforts focus on minimizing runtime overhead by terminating the program early and reducing instrumentation requirements.

\textbf{Pruning-Guided Directed Fuzzing.}
Timely termination of the execution of unreachable inputs can reduce runtime overhead. Beacon \cite{huang_beacon_2022} adopts lightweight static analysis to prune infeasible paths based on control-flow reachability and path-condition satisfiability. SieveFuzz \cite{srivastava_one_2022} restricts execution within reachable regions by identifying unreachable boundaries and immediately terminating execution upon reaching the region boundaries.

\textbf{Selective Instrumentation.}
DGF acquires feedback by instrumenting the target program, albeit at the cost of additional runtime overhead and a slowdown in the fuzzing process. Existing approaches mitigate this overhead through selective instrumentation. SelectFuzz \cite{luo_selectfuzz_2023} selectively instruments and explores only relevant code for fuzzing targets, including path-divergent and data-dependent code. DAFL \cite{kim_dafl_2023} obtains coverage feedback solely from segments of the target program that are pertinent to the target location, identified through thin slicing performed on the DUG. DAFL exclusively performs instrumentation on all basic blocks within functions of the sliced DUG. While these selective instrumentation strategies reduce instrumentations, inaccurate analysis can result in insufficient feedback collection. To address this issue, our tool employs a multidimensional selective instrumentation strategy that precisely instruments necessary basic blocks without sacrificing feedback details.

% 总结
\section{Conclusion}
\label{sec:conclusion}
In this study, we present \tool, a novel directed grey-box fuzzing framework leveraging hybrid runtime feedback, addressing three critical challenges with tailored solutions: our static analyzer uses a backward-stepping algorithm to compute basic block-level control-flow distances on a virtual ICFG for optimized seed prioritization; we introduce value-flow influence and define a value-flow influence score fitness metric to guide systematic exploration of target state spaces; and a selective instrumentation strategy mitigates runtime overhead from hybrid feedback by targeting critical code sections. Empirical evaluations on real-world vulnerabilities show that \tool outperforms state-of-the-art fuzzers like AFL, AFLGo, WindRanger, DAFL, and Beacon in crash reproduction speed and code coverage efficiency.

% 致谢
% \section*{Acknowledgment}
% The authors thank the anonymous reviewers for their helpful suggestions and comments.
% The preferred spelling of the word ``acknowledgment'' in America is without 
% an ``e'' after the ``g''. Avoid the stilted expression ``one of us (R. B. 
% G.) thanks $\ldots$''. Instead, try ``R. B. G. thanks$\ldots$''. Put sponsor 
% acknowledgments in the unnumbered footnote on the first page.

% 引用
\bibliographystyle{bib/IEEEtran}
% \bibliography{bib/abrv,bib/mybib,bib/references}
\bibliography{bib/mybib}

% 附录
% \input{appendix}

\end{document}